\documentstyle[12pt]{article}
\newcommand{\be}{\begin{eqnarray}}
\newcommand{\en}{\end{eqnarray}}
\newcommand{\ka}{\kappa}
\newcommand{\ga}{\gamma}

\begin{document}
\begin{titlepage}
\begin{flushright}
 EFI 93-71

\end{flushright}

\begin{center}
\vskip 0.3truein

{\large{\bf{ASYMPTOTIC LIMITS AND SUM RULES}}}
\vskip 0.15truein

{\large{\bf{FOR GAUGE FIELD PROPAGATORS}}}
\footnote{Work supported in part by the National
Science Foundation, Grant PHY 91-23780}
\vskip 0.6truein

{\large{ Reinhard Oehme and Wentao Xu}}
\vskip 0.2truein

Enrico Fermi Institute and Department of Physics

University of Chicago

Chicago, Illinois 60637, USA
\end{center}
\vskip 0.5truein

\centerline{Abstract}

For gauge field propagators, the asymptotic behavior
is obtained in all directions of the complex $k^2$-plane,
and for general, linear, covariant gauges. Asymptotically
free theories are considered. Except for coefficients,
the functional form of the leading asymptotic terms
is gauge-independent. Exponents are determined exactly
by one-loop expressions. Sum rules are derived, which
generalize the superconvergence relations obtained
in the Landau gauge.
\end{titlepage}
\newpage

\bigskip
\setcounter{equation}{0}

Interesting sum rules for the structure functions of propagators can be
derived on the basis of their analytic properties, together with the
asymptotic behavior for large momenta as obtained with the help of the
renormalization group.  For systems with a limited number of matter
fields, one obtains superconvergence relations for the gauge field
propagator in the Landau  gauge \cite{OZ,Osup}.
These relations are of interest
in connection with the problem of confinement \cite{Ocnf,Npisa}.
Other results are
dipole representations, and information about the discontinuity of the
gauge field structure functions. They indicate the existence of an
approximately linear quark--antiquark potential \cite{Np,Op}, and are
important for understanding the structure of the theory in the state
space with indefinite metric \cite{Omod}.

It is the purpose of this note to present results for the gauge field
propagator in general, covariant, linear gauges.  We obtain the
asymptotic terms for large momenta, and for all directions in the
complex $k^2$--plane.  Sum rules are derived, which generalize the
superconvergence relations of the Landau gauge.  An important aspect of
our results is the gauge--independence of the functional form of the
essential asymptotic terms.  Only the coefficients of these terms depend
upon the gauge parameter.

\bigskip

We consider a non--Abelian gauge theory like QCD, with the gauge--fixing
part of the Lagrangian given by $-B\cdot (\partial_\mu A^\mu) +
\frac{\alpha}{2} B\cdot B$, where $B$ is the usual auxiliary field.  For
$\alpha \neq 0$, $B$ can be eliminated by $\alpha B = (\partial_\mu
A^\mu)$.  In order to define the structure function of the transverse
gauge field propagator, we write
\begin{eqnarray}
&~&\int dx e^{ikx}   \langle 0 | T A^{\mu \nu}_{a}(x)
A^{\varrho\sigma}_{b}(0) |0\rangle~~ =~~
- i \delta_{ab} D (k^2 + i0)\cr
&~&~~~~~~~\times\left( k^\mu k^\varrho g^{\nu \sigma} -
k^\mu k^\sigma g^{\nu \varrho}
+ k^\nu k^\sigma g^{\mu\varrho} - k^\nu k^\varrho g^{\mu\sigma}\right)
\end{eqnarray}
with $A^{\mu \nu}\equiv\partial^\mu A^\nu - \partial^\nu A^\mu$.
We assume the general postulates of covariant gauge theories.
Important are Lorentz covariance and simple spectral conditions,
as formulated in references like \cite{KO,Sc} for
state spaces with indefinite metric. Exact
Green's functions should be connected with the formal perturbation
series in the coupling parameter $g$ for $g^2\to + 0$, at least as far as
the first few terms are concerned.
Topological aspects of the gauge theory are not expected to influence the
asymptotic behavior we consider here.

As a consequence of Lorentz covariance, and the spectral conditions
mentioned above, it follows that the function $D (k^2 + i0)$ is
the boundary value of an analytic function, which is regular in the  cut
$k^2$--plane, with a branch line along the positive real axis.
In contrast to the situation for higher Green's functions
\cite{BOT} , explicite use of local commutativity is not
required for the two-point functions \cite{OZ}. Using
renormalization group methods, together with analyticity, we obtain the
asymptotic behavior for $k^2\to \infty$ in {\it all directions} of the
complex plane.  We present first the essential leading terms, leaving
derivation and details for later.

For the analytic structure function $D (k^2)$,
we find for $k^2\to\infty$ in all directions:
\begin{eqnarray}
-k^2 D (k^2,\ka^2,g,\alpha)&\simeq& \frac{\alpha}{\alpha_0}
 + C_R (g^2, \alpha) \left(-\beta_0 \ln
\frac{k^2}{\ka^2}\right)^{-\gamma_{00}/\beta_0} + \cdots\cr
&~& ~~~~~~+\frac{\alpha}{\alpha_0} C_1 \left(-\beta_0 \ln
\frac{k^2}{\ka^2}\right)^{-1} + \cdots
\label{dk}
\end{eqnarray}
The corresponding asymptotic terms for the discontinuity along the
positive, real $k^2$--axis are then given by
\begin{eqnarray}
-k^2 \rho (k^2,\ka^2,g,\alpha) & \simeq& \frac{\gamma_{0 0}}{\beta_0} C_R
(g^2,\alpha) \left(-\beta_0 \ln \frac{k^2}{\vert
\ka^2\vert}\right)^{-\gamma_{0 0}/\beta_0 - 1} + \cdots\cr
&~&~~~~~~ +\frac{\alpha}{\alpha_0} C_1 \left(-\beta_0 \ln \frac{k^2}{\vert
\ka^2\vert}\right)^{-2} + \cdots
\label{rk}
\end{eqnarray}
The parameters and their limitations are as follows:    The function $D$
is normalized at the real point $k^2 = \ka^2 < 0$, where
\be
-k^2 D (k^2,\ka^2,g,\alpha)\vert_{k^2 = \ka^2} = 1.
\label{norm}
\en
The anomalous dimension of the gauge field is given by $\gamma
(g^2,\alpha) = \gamma_0 (\alpha) g^2 + \gamma_1 (\alpha) g^4
+\cdots$, $\gamma_0 (\alpha) = \gamma_{0 0} + \alpha \gamma_{01},
  \gamma_1 (\alpha) = \gamma_{10} +
\alpha \gamma_{11} + \alpha^2\gamma_{12}$,
etc., and $\alpha_0 \equiv - \gamma_{0 0}/\gamma_{01}$.  The
renormalization group function is $\beta (g^2) = \beta_0 g^4 +
\beta_1 g^6 + \cdots$.  For QCD, we have $\gamma_{0 0} = - (16
\pi^{2})^{-1} (\frac{13}{2} -\frac{ 2}{3} N_F),
 \gamma_{01} = (16 \pi^{2})^{-1}\frac {3}{2},$
$\beta_0 = - (16 \pi^{2})^{-1} (11 - \frac{2}{3}N_F),
 N_F$ = number of flavors.
We assume $\beta_0 < 0$ corresponding to asymptotic freedom.
Consequently, the exponent $\gamma_{00}/\beta_0$ in eqs. (2) and (3)
varies from $13/22$ for $N_F = 0$ to 1/10 for $N_F = 9$, and from
$-1/16$ for $N_F = 10$ to $-15/2$ for $N_F = 16$.  We have $0 <
\gamma_{00}/\beta_0 < 1$ for $N_F\le 9$ and
$\gamma_{00}/\beta_0 < 0$ for $10 \le N_F \le 16$;
 for $\gamma_{00}/\beta_0 = 1$, our relations
(\ref{dk}) and (\ref{rk}) would require modifications.

The essential asymptotic term in eqs.(\ref{dk})
and (\ref{rk}) is the one with the
coefficient $C_{R} (g^2,\alpha)$, which is not identically zero,
although there may be zero surfaces $\alpha = \alpha_0 (g^2)$.  In case
$C_{R}$ should vanish, the $C_1$ -- term becomes relevant.  Its
coefficient is given by
\begin{eqnarray}
C_1 &=& \frac{\gamma_1 (\alpha_0)}{\beta_0 - \gamma_{00}}\ \hbox{~~~~for}\ 0 <
\frac{\gamma_{00}}{\beta_0} < 1,\cr
C_1 &=& \frac{\gamma_{00} + \alpha_0 \gamma_{11}}{\beta_0 - \gamma_{00}} \
\hbox{~~~~for}\ \frac{\gamma_{00}}{\beta_0} < 0.
\end{eqnarray}
For $\alpha = 0$, the coefficient $C_{R}$ is positive and given by
\be
C_R (g^2,0) &=& (g^2)^{-\gamma_{00}/\beta_0} \hbox{exp}\int^0_{g^2}
dx  \tau_0 (x), \cr
\tau_0 (x) &\equiv& \frac {\gamma (x,0)}{\beta (x)} -
\frac {\gamma_{00}}{\beta_0 x }.
\label{CR0}
\en
Eq. (\ref{CR0}) follows from the exact
solution for $\alpha = 0$, with the
normalization $-{\ka}^2D({\ka}^2, {\ka}^2, g, 0) = 1$ \cite{OZ,Osup}.
This solution can be written in the form
\be
-k^2D(k^2, \ka^2, g, 0) ~~=~~ \left( \frac{\bar g^2}
{g^2}\right)^{\gamma_{00}/\beta_0}
\hbox{exp} \int^{\bar g^2}_{g^2} dx \tau_0 (x).
\label{D0}
\en
Here ${\bar g^2} = {\bar g^2}(u,g)$, $u = \vert \frac{k^2}{\ka^2}
\vert$ is the effective gauge coupling, with the properties
\be
\ln u = \int^{\bar g^{2} (u,g)}_{g^2} dx \beta^{-1} (x), \cr
\bar g^2 (u,g) \simeq  1/({-\beta_0 \ln u }) + \cdots
 \label{geff}
\en
for $u\to\infty$, in the case of asymptotic freedom with $\beta_0 < 0$.

In general, $C_{R} (g^2,\alpha)$ satisfies a partial differential
equation.  In an approximation, where only terms linear in $\alpha$
are kept in the expression for the anomalous dimension $\gamma
(g^2,\alpha)$, we obtain
\be
C_{R} (g^2,\alpha)& = & (g^2)^{-\gamma_{00}/\beta_0} \
\hbox{exp}\left( \int^0_{g^2} dx \tau_0 (x)\right)\cr
&\times&  \left\{ 1 - \frac{\alpha}{\alpha_0} + \frac{\alpha}{\alpha_0}
(g^2)^{\gamma_{00}/\beta_0} \int^0_{g^2} dx x^{-\gamma_{00}/\beta_0}
f(x,g^2)\right\}.
\label{CR1}
\en
Here
\be
f (x,g^2)& =& \{ \tau_0 (x) + \alpha_0 \tau_1 (x)\}
\hbox{exp}\int^{g^2}_x dy \tau_0 (y),\cr
\tau_1(x)& =& {\bar \gamma}_1 (x)/\beta (x) - \gamma_{01}/\beta_0 x,\cr
\gamma (x,\alpha)& =& {\bar \gamma}_0 (x) + \alpha
{\bar \gamma}_1 (x) + O(\alpha^2).
\label{tau1}
\en
 In eq. (\ref{CR1}), the normalization
for the full solution in the
 the $\alpha$--linear approximation has been used
in order to fix an otherwise
 undetermined coefficient of $C_{R}$.
This approximation consists of replacing the general
anomalous dimension $\gamma (g^2 ,\alpha )$ by the
$\alpha $-linear form ${\bar \gamma}_0 (g^2 ) +
 \alpha {\bar \gamma}_1 (g^2 )$.
The corresponding solution is given by
\be
-k^2D(k^2, \ka^2, g, \alpha)& = & \hbox{exp} \left (
\int^{{\bar g}^2}_{g^2} dx
\frac{{\bar \ga}_0 (x)}{\beta (x)}\right )\cr
&\times & \left\{ 1 + \alpha \int^{{\bar g}^2}_{g^2} dx
\frac {{\bar \gamma}_1 (x)}{\beta (x)}
\hbox{exp} \left ( - \int^{x}_{g^2} dy \frac {{\bar \ga}_0 (y)}
{\beta (y)} \right ) \right \},
\label{Dlin}
\en
with the notation as defined in eqs. (\ref{geff}) and
(\ref{tau1}).

 An  important aspect of the leading asymptotic terms for
$$
-k^2 D (k^2, \ka^2, g, \alpha) -
 \frac{\alpha}{\alpha_0}~~~ \hbox{and}~~~ -k^2 \rho (k^2, \ka^2, g, \alpha)
$$
 is their independence
 of the gauge parameter $\alpha$, except
for the coefficient $C_{R} (g^2 ,\alpha)$.
 In addition, the exponents of the logarithms in eqs. (\ref{dk})
 and (\ref{rk}) are completely and exactly determined by
 one--loop coefficients of the anomalous dimension $\gamma (g^2, 0)$ and
of the $\beta$--function.
\bigskip
\bigskip

 In view of the asymptotic behavior of $D(k^2,\ka^2,g,\alpha)$ as given in
 eq. (2), which is valid for all directions in the $k^2$--plane, we can
 write the usual unsubstracted dispersion representation
 \be
D (k^2,\ka^2,g,\alpha) &=& \int^\infty_{-0} dk'^{2}  \frac{\rho (k'^{2},
 \ka^2, g, \alpha)}{k'^{2} - k^2}
\label{dispr}
 \en
 We also have sufficient boundedness for
 the discontinuity $\rho$ in order to
 write a dipole representation
 \begin{eqnarray}
 D (k^2,\ka^2,g,\alpha) &=& \int^\infty_{-0} dk'^{2} \frac{\sigma
 (k'^{2},\ka^2,g,\alpha)}{(k'^{2} - k^2)^2},\cr
 \sigma (k^2,\ka^2,g,\alpha) &=& \int^{k^2}_{-0} dk'^{2} \rho
 (k'^{2},\ka^2,g,\alpha).
 \end{eqnarray}
 For $\alpha = 0$, the dipole representation has been discussed
 in refs. \cite{Op} and \cite{Np} in connection with an approximately linear
 quark--antiquark potential.

 Of particular interest is the situation for $0<\gamma_{00}/\beta_0 < 1$,
 corresponding to $N_F\le 9$ in QCD.  There, the function $D +
 \frac{\alpha}{\alpha_0} k^{-2}$ vanishes faster than $k^{-2}$ for
 $k^2\to\infty$, and hence we have the sum rule
 \be
\int^\infty_{-0} dk^2\rho (k^2,\ka^2,g,\alpha) &=&
 \frac{\alpha}{\alpha_0}.
\label{sup}
 \en
 This is the generalization of the superconvergence relation
\cite{OZ,Osup}
 $$\int^\infty_{-0} d k^2 \rho (k^2,\ka^2,g,0) ~~=~~ 0,
 $$
 which was obtained previously in the Landau gauge.  The relation
(\ref{sup}) expresses the fact that the coefficient of the
asymptotic term proportional to $k^{-2}$ in the representation
(\ref{dispr}) is given by $-\alpha / \alpha_0 $. It
is {\it not} valid for $\gamma_{00}/\beta_0 < 0$.
The distribution aspects of sum rules like eq. (\ref{sup}),
and of the related dispersion representations, have been discussed
in refs. \cite{OZ,Osup}.

\bigskip
\bigskip

 In order to derive the asymptotic properties
of the structure function $D$, we consider
 the renormalization group equation for
the dimensionless function $R (k^2/\ka^2, g,
 \alpha)\equiv - k^2 D(k^2,\ka^2,g,\alpha)$.
 We obtain
 \be
R\left( \frac{k^2}{\ka^2}, g, \alpha\right) &=& R \left(
 \frac{\ka'^{2}}{\ka^2}, g, \alpha\right) R \left( \frac{k^2}{\ka'^{2}}, \bar
 g, \bar\alpha\right),
\label{ren}
\en
where we have used the relation
$$R^{-1} (\ka'^{2}/\ka^2,g,\alpha) = Z_3 (\ka'^{2}/\ka^2,g,\alpha),$$
which follows from the normalization condition (\ref{norm}):
$R(1,g,\alpha) = 1$ at $k^2 = \ka^2 < 0$ , and where
$Z_3$ is the square of the conventional renormalization
factor for the gauge field. Further, in eq. (\ref{ren}),
$\bar g = \bar g (\frac{\ka'^{2}}{\ka^2}, g)$
 is the effective (running) gauge
 coupling parameter, and $\bar \alpha$ is given by
$\bar\alpha = \alpha R^{-1} \left(
 \frac{\ka'^{2}}{\ka^2}, g, \alpha\right)$.
  We use a mass
 independent renormalization scheme \cite{WHC}, which is appropriate for the
 study of asymptotic limits.

 Let us first consider the limit $k^2\to - \infty$ along the negative, real
 $k^2$--axis, where $R(k^2/\ka^2, g, \alpha)$ is analytic and real.  We set
 $ u=\vert k^2/\ka^2 \vert$ and define the function
  $R(\bar {g}^{2}; g^2,\alpha)\equiv R(u,g,\alpha)$, with $\bar g (u,g)$
 being the effective coupling defined in eq. (\ref{geff}).
 From eq. (\ref{ren}), we then obtain the differential equation
 \be
 \beta (\bar g^2) \frac{\partial R(\bar g^2;
 g^2,\alpha)}{\partial\bar g^2} = \gamma (\bar g^2, \bar\alpha) R (\bar
 g^2; g^2,\alpha),\nonumber \\
 \bar\alpha = \bar\alpha (\bar g^{2}; g^2,
\alpha)\equiv\alpha R^{-1} (\bar g^2;
 g^2,\alpha).
\label{dir}
 \en
 For $\alpha\not= 0$, it is more convenient to work with the equation
 \be
\beta (\bar g^2) \frac{\partial\bar\alpha}{\partial\bar g^2} = -
 \bar\alpha \gamma (\bar g^2,\bar\alpha),
\label{dia}
 \en
 with $\bar\alpha$ as defined in eq. (\ref{dir}).
 From the general solution (\ref{D0}) of
 eq. (\ref{dir}) for $\alpha = 0$,
 as well as the solution (\ref{Dlin})
 for $\alpha\not= 0$ in the $\alpha$--linear
 approximation, we know that R and $\bar\alpha$ have a branch
 point as a function of $\bar g^2$ at $\bar g^2 = 0$,
 which is of the form
 $(\bar g^2)^\xi, \xi \equiv \gamma_{00}/\beta_0$.
 For QCD, we have
$\xi = \frac{\frac{13}{2}-\frac{2}{3}N_F}{11-\frac{2}{3} N_F}$.
We see that $\vert\xi\vert = n/d$ is rational, with $n$ and $d$
 being relative primes.  It is then convenient to uniformize the
 algebraic branch point by introducing $x=(\bar g^2)^{1/d}$ as a
 uniformization variable.
\bigskip

 We consider first the case $0 < \xi < 1$, corresponding
 to $N_F \leq 9$ for QCD.  Here it is convenient to use
 eq. (\ref{dir}).  We define $y (x) = (\bar\alpha - \alpha_0) x^{-n}$, and
 obtain the differential equation
 \be
\frac{dy}{dx} = \frac{n}{\alpha_0} x^{n-1} y^2 - d~ x^{d-n-1} (\alpha_0
 + x^n y) \phi (x^d,\alpha_0  + x^n y),
\label{diy}
 \en
 where
 \begin{eqnarray}
 \phi (g^2,\alpha)&  \equiv &\frac{\gamma (g^2,\alpha)}{\beta (g^2)} -
 \frac{\gamma_0 (\alpha)}{\beta_0 g^2} = \phi_0 (\alpha) + g^2\phi_1
 (\alpha) + \cdots\cr
 \phi_0 (\alpha)& =& \frac{\gamma_1 (\alpha)}{\beta_0} -
 \frac{\beta_1}{\beta_0^2} \gamma_0 (\alpha),~~~\hbox{etc.},
\label{phi}
\end{eqnarray}
 with the definitions given below eq. (\ref{norm}).

 In an appropriate finite domain including $g^2=0$, and excluding
 possible, nontrivial fixed points corresponding to zeroes of $\beta
 (g^2)$, it is reasonable to assume that $\phi (g^2,\alpha)$ is
 continuously differentiable.  As far as $\beta (g^2)$ and $\gamma
 (g^2,\alpha)$ are represented by power series expansions for $g^2\to +
 0, \phi (g^2,\alpha)$ is also a power series in $g^2$ and $\alpha$.
 Under these circumstances, the r.h.s. of
 eq. (\ref{diy}) satisfies the Lipschitz
 condition for $x=0$, and we have exactly one solution through every
 point $x=0, y=C$.  In as far as the r.h.s.
 of eq. (\ref{diy}) is also a power
 series, we obtain the solution in the form of a series:
 \begin{eqnarray}
 y(x) &=& C + \frac{C^2}{\alpha_0} x^n
  +C \left( \frac{\xi + 1}{\xi - 1} \phi_0 (\alpha_0) - \alpha_0 \phi'_0
 (\alpha_0)\right) x^d + \cdots \cr
 &~&~~~~~~+ \frac{\alpha_0}{\xi - 1} \phi_0 (\alpha_0) x^{d-n} + \cdots,
\label{yx}
 \end{eqnarray}
where we have separated the terms proportional to $C$.

 For the asymptotic expansion of $R (\bar g^2; g^2,\alpha)$ for $\bar
 g^2\to + 0$, eq. (\ref{yx}) implies
 \begin{eqnarray}
 R(\bar g^2; g^2,\alpha) &\simeq& \frac{\alpha}{\alpha_0} + C_{R} (\bar
 g^2)^\xi +
 C_{R} \beta^{-1}_0 \left(\gamma_{10} - \gamma_{12}\alpha_0^2 -
 \frac{\beta_1}{\beta_0}\gamma_{00}\right) (\bar g^2)^{\xi + 1} +\cdots\cr
 &~&~~~~~~+ \frac{\alpha}{\alpha_0} \frac{\gamma_1 (\alpha_0)}{\beta_0}
 \frac{1}{1-\xi} \bar g^2 + \cdots,
\label{rgg}
 \end{eqnarray}
 with $C_{R} = - C\alpha/\alpha^2_0$, and $0 < \xi < 1$.  This formula
 is also valid for $C_{R} = 0$. The term
 proportional to $C^2$ in eq.(\ref{yx}) cancels in the inversion
 leading to eq.(\ref{rgg}).

 For $\xi < 0$, corresponding to $10 \leq N_F \leq 16$ for QCD,
 it is more convenient to use eq. (\ref{dir}).  With $\xi = -
 \frac{n}{d}$, $n$ and $d$ being positive integers which are
 relative primes, we uniformize the branch point at $\bar g^2 = 0$ by
introducing again a new variable $x$ so that
$\bar g^2 = x^d$. Then we define $z(x) = x^n R(x^d; g^2, \alpha)$,
and obtain the differential equation
\be
 \frac{dz}{dx} = \alpha_0\alpha nx^{n-1} + d~ x^{d-1} z \phi (x^d,
 \frac{\alpha x^n}{z})
 \label{diz}
\en
 As long as $x^n z^{-1}$ remains bounded around $x=0$, we have again the
 Lipschitz condition satisfied and obtain the power series solution
 \begin{eqnarray}
 z(x) &=& C_{R} + C_{R}\phi_0 (0) x^d + \cdots \cr
&+& \frac{\alpha}{\alpha_0} x^n
 + \frac{\alpha}{\alpha_0} \frac{d}{d+n}
 (\phi_0 (0) + \alpha_0 \phi'_0 (0)) x^{d+n}  + \cdots ,
\label{zx}
 \end{eqnarray}
 with $C_{R}\not= 0$.  In terms of
 $R(\bar g^2; g^2, \alpha)$, eq. (\ref{zx})
 leads to the asymptotic expansion for $\xi < 0$ and $\bar g^2\to + 0$:
 \begin{eqnarray}
 R (\bar g^2; g^2, \alpha) &\simeq&C_{R} (\bar g^2)^\xi + C_{R}
 \frac{1}{\beta_0} (\gamma_{10} - \frac{\beta_1}{\beta_0} \gamma_{00})
 (\bar g^2)^{\xi + 1} +\cdots\cr
 &~&~~~~~~+ \frac{\alpha}{\alpha_0} +
 \frac{\alpha}{\alpha_0} \frac{(\gamma_{10} +
 \alpha_0 \gamma_{11})}{\beta_0 (1-\xi)}  \bar g^2 + \cdots
\label{rgz}
 \end{eqnarray}
 For $C_{R} = 0$, we cannot use eq. (\ref{diz}),
 but obtain the asymptotic expression directly from eqs.
 (\ref{yx}) and (\ref{rgg}) :
\be
R(\bar g^2; g^2, \alpha) \simeq \frac{\alpha}{\alpha_0} +
 \frac{\alpha}{\alpha_0} \frac{\ga_1 (\alpha_0)}{\beta_0}
 \frac{1}{1 + \vert \xi \vert}  \bar g^2 + \cdots.
\en
This relation corresponds to eq. (\ref{rgg}) with
 $C_{R} = 0$ and $\xi < 0$.

 With eqs. (\ref{rgg}) and (\ref {rgz}),  we have
obtained the asymptotic expressions for
 $R(\frac{k^2}{\ka^2}, g, \alpha)$
 in the limit $k^2\to - \infty$ along the real axis,
provided we can use
 $\bar g^2 (u,g)\simeq -1/(\beta_0 \ln u) + \cdots$.  It remains to
 consider the limit $k^2\to\infty$ in all directions of the complex
 $k^2$--plane.  We return to eq. (\ref{ren}).
 Setting $\ka^{\prime 2} = - \vert
 k^2\vert$, we find, with $\ka^2 < 0$ and $k^2 = - \vert k^2\vert
 e^{i\varphi}$ for all $ \vert\varphi\vert \leq \pi$:
 \be
R \left( \frac{k^2}{\ka^2},g,\alpha\right) =
 R \left(\left\vert \frac{k^2}{\ka^2}
 \right\vert, g,\alpha\right) R(e^{i\varphi},\bar g, \bar\alpha).
 \label{rep}
 \en
 Here $\bar g = \bar g (\vert\frac{k^2}{\ka^2}\vert,g)$ and $\bar\alpha =
 \alpha R^{-1} (\vert\frac{k^2}{\ka^2}\vert,g,\alpha)$.
  For $\beta_0 < 0$, the effective coupling $\bar g^2$
 vanishes for $\vert\frac{k^2}{\ka^2}\vert\to\infty$, and  $\bar\alpha$
 remains bounded in this limit, as may be seen from eqs. (\ref{rgg})
 and (\ref{rgz}).  Because $R$ is analytic in the cut
 complex $k^2$--plane, we can then use the perturbation expansion for the
 structure function,
 \be
R \left( \frac{k^2}{\ka^2},g,\alpha \right) \simeq 1 + g^2
 \gamma_0 (\alpha ) ln \left ( \frac{k^2}{\ka^2} \right ) + O(g^4),
 \en
 and write for $\bar g^2\to + 0$:
 \be
R (e^{i\varphi},\bar g,\bar\alpha)\simeq 1+\bar g^2
 \gamma_0 (\bar \alpha) i\varphi + O({\bar g}^4) .
\label{rph}
 \en
 Eq. (\ref{rep}) expresses
the asymptotic limit for $k^2\to\infty$ in
 all directions in terms of the limit along the negative real
 $k^2$--axis.  With eqs. (\ref{rep}), (\ref{rph}),
 (\ref{rgg}) and (\ref{rgz}), we finally obtain the
 limits given in eqs. (\ref{dk}) and (\ref{rk}).

 A priori, the coefficients $C$ or $C_{R}$
 appearing in the solutions of
 the nonlinear, ordinary differential equations are undetermined
 constants. However, because of the normalization condition $R(g^2;
 g^2,\alpha) =1$ or $\bar\alpha (g^2; g^2,\alpha) = \alpha$, the
 coefficients become functions of $g^2$ and $\alpha$, satisfying
 partial differential equations in these variables.  For $C_{R}
 (g^2,\alpha)$, we find the equation
 \begin{eqnarray}
 C_{R} (g^2,\alpha) &=& R(g'^{2}; g^2,\alpha) C_{R} (g'^{2},\alpha')\cr
 \alpha' &=& \alpha R^{-1} (g'^{2}; g^2,\alpha),
 \end{eqnarray}
 and the corresponding differential equation is:
\be
 \beta (g^2) \frac{\partial C_{R}}{\partial g^2} = \alpha \gamma
 (g^2,\alpha) \frac{\partial C_{R}}{\partial\alpha} - \gamma
 (g^2,\alpha) C_{R}.
\label{cg}
\en

 For $\alpha = 0$, and in the $\alpha$--linear approximation, we have
 given $C_{R}$ in eqs. (\ref{CR0}) and (\ref{CR1}), which satisfy
 eq. (\ref{cg}) with $\alpha = 0$ and $\ga (g^2, \alpha) =
{\bar \ga}_0 (g^2) + \alpha {\bar \ga}_1 (g^2)$ respectively.
\bigskip
\bigskip

 Several of the results presented in
 this paper have been obtained by one
 of us (R.O.) in collaboration with W. Zimmermann \cite{OZu}.
 It is a pleasure to thank Wolfhart Zimmermann for his
 contribution and for many discussions.

\end{document}